\documentclass[11pt]{article}
\usepackage{epsfig}
\usepackage[usenames]{color}
\usepackage{amsmath}
\usepackage{amsfonts}
\usepackage{amssymb}
\newtheorem{proposition}{Proposition}[section]
\newcommand{\bpr}{\begin{proposition}}
\newcommand{\epr}{\end{proposition}}

\newcounter{Roman}
\addtocounter{Roman}{1}

\newcommand{\beq}{\begin{equation}}
\newcommand{\eeq}{\end{equation}}
\newcommand{\bea}{\begin{eqnarray}}
\newcommand{\eea}{\end{eqnarray}}
\newcounter{saveeqn}

\newcommand{\ssc}{\scriptscriptstyle}

\newcommand{\hnu}{\hat{\nu}}
\newcommand{\hmu}{\hat{\mu}}
\newcommand{\hkappa}{\hat{\kappa}}
\newcommand{\tr}{{\rm tr}}

\newcommand{\re}{{\rm Re}}

\newcommand{\vev}[1]{\Big\langle #1 \Big\rangle}
\newcommand{\bpsi}{\bar{\psi}}

\newcommand{\N}{{\cal N}}
\newcommand{\R}{{\cal R}}

\input epsf
\begin{document}  

\begin{center}{\Large\bf  Absence of chiral symmetry breaking in multi-flavor strongly coupled lattice gauge theories }\\[2cm] 
{E. T. Tomboulis\footnote{\sf e-mail: tomboulis@physics.ucla.edu}
}\\
{\em Department of Physics and Astronomy, UCLA, Los Angeles, 
CA 90095-1547}
\end{center}
\vspace{1cm}

\begin{center}{\Large\bf Abstract}
\end{center} 
We consider lattice gauge theories at strong coupling with gauge group $U(N_C)$, or $SU(N_C)$  restricted to the meson sector, 
and coupled to $N_F$ flavors of fundamental representation staggered fermions. We study the formation of a chiral condensate by means of resummation of a hopping expansion. Different classes of graphs become dominant as the parameter $(N_F/N_C)$ is varied. By performing graph resummation we obtain an 
equation for determining the condensate as a function of $(N_F/N_C)$ and mass $m$. For values of 
$(N_F/N_C)$ below a critical value one reproduces the well-known result of the existence of a 
non-vanishing condensate solution in the $m=0$ limit. Above the critical $(N_F/N_C)$ value, however, no such solution exists, its abrupt disappearance indicating a first order transition to a 
chirally symmetric phase with composite (colorless) excitation  spectrum. 
\vfill
\pagebreak

\section{Introduction} 

Chiral symmetry breaking in QCD is one of the cornerstones of modern particle physics. The existence of a non-vanishing chiral condensate at $T=0$ has of course been established and studied in any number of Monte Carlo investigations over several decades \cite{R}. The phenomenon can be nicely demonstrated analytically in lattice QCD at strong coupling as was done  for large number of colors $N_C$ long ago \cite{BBEG} - \cite{KS}. 

Conventional wisdom seems to have been that chiral symmetry will eventually be broken by taking the coupling sufficiently strong regardless of the number of fermion flavors. The reason one might suspect this is not correct is discernible in the very same analytic 
demonstrations of chiral symmetry breaking at strong coupling: as the number of flavors becomes 
large compared to the number of colors, contributions in the fermion hopping expansion that used to be subdominant grow to be dominant, and the original result of a non-vanishing chiral condensate is no longer necessarily valid. 

In this paper we use the same framework of lattice gauge theory at strong coupling and 
gauge group $U(N_C)$, or $SU(N_C)$ restricted to the mesonic sector, 
but with any number $N_F$ of flavors of fundamental representation staggered fermions, and  employ a hopping expansion to study the formation of a chiral condensate. 
To see whether a condensate forms in the chiral limit $m \to 0$ requires resummation of infinite classes of graphs allowing one to obtain an equation for the condensate. One then finds that the outcome depends on the parameter $(N_F/N_C)$. There is a critical value of this parameter below which the usual non-vanishing condensate forms. 
Above this critical value, however, the non-vanishing condensate solution abruptly ceases to exists 
as a physical solution (turns complex), which is the usual sign of a first order phase transition. 

Such a result has in fact already been demonstrated for the case $N_C=3$ in a recent Monte Carlo study \cite{deFetal}. These authors also provide evidence that the chiral phase resulting above the 
critical number of flavors possesses a massless spectrum.  

It should be noted that it is perfectly straightforward to set up a $1/N_F$ expansion in the continuum theory by, as usual, keeping a 't Hooft coupling $\lambda^2=N_Fg^2$ fixed and 
appropriately rescaling the gauge field and fermion sources. Such an expansion \cite{F0}, however, is   not useful for the problem addressed here, i.e., the strong coupling regime, as it is incapable of providing even a hint of the actual spectrum. For that one has to use the nonperturbative lattice formulation, to which this continuum expansion set up cannot be carried over, and a rather more involved procedure is needed to deal with the $N_F$ versus $N_C$ dependence.

Existence of strongly coupled gauge systems with composite (colorless) spectrum and chiral symmetry is of great physical and theoretical interest on many fronts. 
Apart from the basic questions of the complete phase diagram in $(\beta, (N_F/N_C), m)$,  
and the relation to possible continuum limits, there is a host of potential applications for such systems both in particle and condensed matter physics (cf remarks in section \ref{Con}). 
A lot of work will be required to adequately explore these various directions. 

The paper is organized as follows. In section \ref{largeNF} we set up the expansion and 
classify graphs according to their dependence in $N_F$, $N_C$ and (inverse) powers in $m$ . 
This allows us to see how to group graphs in classes to be resummed so that the small $m$ limit may eventually be considered. The resummation is then done in section \ref{resumeq} and the equation for the condensate obtained. Its solutions are considered in section \ref{noCSB}. Section \ref{Con} contains our concluding remarks.

\section{Hopping expansion and $(N_F/N_C)$ graphology  \label{largeNF}}
We work on a euclidean hypercubic d-dimensional lattice 
with lattice sites denoted by their lattice coordinates $x=(x^\mu)$,  and lattice unit vectors in the $\mu$-th direction by $\hmu$. We use standard lattice gauge theory notations and conventions. 
The gauge field bond variable $U_b$ on bond $b=(x,\hmu)$ is more explicitly denoted by $U_\mu(x)$, and the fermion fields on site $x$ by $\bpsi(x)$ and $\psi(x)$.

The lattice action for massless fermions is given by: 
\beq
S_F= 
\sum_{b=(x,\mu)}{1\over 2} \,\left[ \bar{\psi}(x) \gamma_\mu(x) U_\mu(x) \psi(x+\hat{\mu}) - \bar{\psi}(x+\hat{\mu}) \gamma_\mu(x) U^\dagger_\mu(x) \psi(x) \right]  \,.
\label{act1}
\eeq
The matrices $\gamma[b]=\gamma_\mu(x)$ defined on each bond $b=(x,\mu)$ 
satisfy $\prod_{b\in \partial p}\gamma[b]=1$ for each plaquette $p$. 
For staggered fermions 
\beq
\gamma_\mu(x) = (-1)^{\sum_{\nu< \mu} x^\nu}  \, ,   \qquad \gamma_1(x) = 1 \, .\label{gamma1} 
\eeq
For naive fermions $\gamma_\mu(x) = \gamma_\mu$ are elements of the 
Euclidean Dirac-Clifford algebra satisfying $\{\gamma^\mu, \gamma^\nu\} = 2 \delta^{\mu\nu}{\bf 1}$ and are hermitian matrices. Naive fermions are of course unitarily equivalent to $2^{d/2}$ copies of staggered fermions. 

We take the fermions in $N_F$ flavors or tastes. For staggered fermions this corresponds 
to $4N_F$ continuum flavors.   
The staggered fermion action in (\ref{act1}) is then invariant under a $U(N_F)\times U(N_F)$ global symmetry corresponding to independent rotations of fermions on even and odd sublattices \cite{F1}. 
This symmetry is referred to as chiral symmetry. 

We take the fermions to transform under the fundamental representation of the gauge group which 
is taken to be $U(N_C)$ or $SU(N_C)$. In the case of $SU(N_C)$, however, we restrict to the meson sector in the hopping expansion below. This is not essential and is done to avoid the technical nuisance of having to also include the baryon sector which would made things unnecessarily more cumbersome. 

We consider the order parameter for chiral symmetry breaking $\bpsi(x)\psi(x)$. 
Its vev is related to the fermion 2-point function (full propagator) 
$G^{a i,bj}(x,y)=\vev{\psi^{a i} (x) \bpsi^{bj}(y)}$ in the limit $x=y$:
\beq
\vev{\bpsi(x) \psi(x)}  =  -\tr \left[ G(x,x) \right] \, .\label{exp1}
\eeq
Here $\tr$ denotes trace over color, flavor, and (if naive fermions are considered) Dirac spinor indices. We will denote by $\tr_{\ssc C}$ and $\tr_{\ssc F}$ 
traces over color ($a$) and flavor ($i$) indices, respectively. Note that 
$\bar{G} (x,x)\equiv \tr_{\ssc C} G(x,x)$ is a gauge invariant quantity. 

To study (\ref{exp1})  we add an external source 
coupled to the operator $\bpsi(x)\psi(x)$, i.e. a mass term, to the action. 
We write the lattice gauge theory action in the presence of the source more concisely in the 
form 
\beq 
S= S_G + S_F = \sum_p \beta\,[1- {1\over N_C}\re \,\tr U_p ] + \sum_{x,y} \bpsi(x){\cal K}_{x,y}(U)\psi(y) \; ,\label{act2} 
\eeq
where 
\beq 
{\cal K}_{x,y}(U) = {\bf M}_{x,y}(U) + {\bf K}_{x,y}  \label{act3}
\eeq
with
\bea
{\bf M}_{x,y}(U) & \equiv & \!\!{1\over 2}\!\!\left[ \gamma_\mu U_\mu(x) \delta_{y,x+\hmu} 
\!-\!
\gamma_\mu U^\dagger_\mu (x-\hmu) \delta_{y,x-\hmu} \right] \label{act3a}\\
{\bf K}_{x,y} & = &  m {\bf 1_{\ssc C}} {\bf 1_{\ssc F}}\,\delta_{x,y} \;. \label{act3b}
\eea
In the strong coupling limit $\beta\to 0$ the plaquette term in (\ref{act2}) is dropped. The 
corrections due to this term can be taken systematically into account within the strong coupling cluster expansion, which, for sufficiently small $\beta$, converges. 

Setting $\beta=0$ in (\ref{act2}) then, $G(x,x)$, in the presence of arbitrary source ${\bf K}$, is given by 
\bea
G(x,x) & = & {1\over \int [DU] \, {\rm Det} {\cal K}(U)}
\int [DU] \,  {\rm Det}{\cal K}(U)\,
{\cal K}^{-1}_{x,x}(U)
\label{exp2a}\\
& = & {
\int [DU] \,  {\rm Det} [{\bf 1 + K^{-1} M}(U)]  \; 
   \left[[{\bf 1 + K^{-1} M}(U)]^{-1} {\bf K}^{-1}\right]_{x,x}
\over \int [DU] \, {\rm Det} [{\bf 1 + K^{-1} M}(U)] }  \,,    \label{exp2b}
\eea
The vev of $\bpsi(x)\psi(x)$ in the presence of the source is then 
obtained from (\ref{exp1}).


 We evaluate (\ref{exp2b}) in the hopping expansion. This amounts to expanding (\ref{exp2b}) treating 
 ${\bf M}$ as the interaction and taking ${\bf K}$ as defining the inverse bare propagator:
 \beq
 {\bf K}^{-1}_{x,y} = m^{-1}{\bf 1_{\ssc C}}{\bf 1_{\ssc F}}\, \delta_{x,y} \,.
 \label{bareG}
 \eeq
Note that  ${\bf K}$ is purely local, whereas ${\bf M}$ has only 
nearest-neighbor non-vanishing elements ${\bf M}_{x, x+\hmu} =  {1\over 2}\gamma_\mu U_\mu(x) $ and 
${\bf M}_{x, x-\hmu} = -{1\over 2}\gamma_\mu U^\dagger_\mu(x-\hmu)$.  The expansion can be shown to converge in the large volume limit for sufficiently large $m$. 
In the absence of the plaquette term, and since ${\bf M}$ is linear in the bond variables $U_b$, integration over the gauge field results in non-vanishing contributions only if at least two $M$ factors with equal number of $U_b$'s and $U_b^\dagger$'s 
occur on each bond $b$. In the case of $SU(N_C)$ equal mod $N_C$ number of $U_b$ and $U_b^\dagger$'s are also allowed but will be excluded by restriction to the mesonic sector as already explained above. 
\begin{figure}[ht]
\includegraphics[width=\columnwidth]{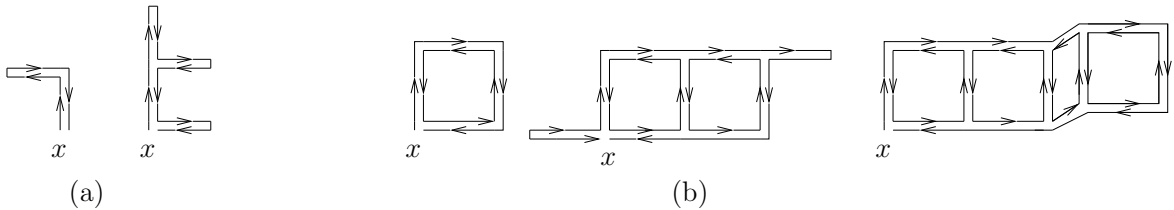}
\caption{(a) Some tree graphs; and (b) some loop graphs attached to site $x$. 
(Arrows indicate $U$, $U^\dagger$'s on bonds.) \label{csscF1}}
\end{figure}

The  expansion of the ${\cal K}^{-1}_{x,x}(U)$ is represented by all paths starting and ending at $x$, whereas that of 
the ${\rm Det} {\cal K}(U)$ by all closed paths \cite{R}, \cite{S}.  Consistent with the above constraint on each bond resulting from the $U$-integrations,  after the cancellation of all disconnected graphs between numerator and denominator the connected graphs giving the expectation 
(\ref{exp2b}) naturally fall into two classes: `tree graphs' and `loop graphs'. 

The tree graphs consist of paths starting and ending at $x$ and 
enclosing zero area (Fig. \ref{csscF1} (a)); they arise entirely from the expansion of 
${\cal K}^{-1}_{x,x}(U)$. 
The loop graphs, such as those depicted in Fig. \ref{csscF1}(b), consist of paths from the expansion of ${\cal K}^{-1}_{x,x}(U)$ and loops from that of ${\rm Det} {\cal K}(U)$ coupled by the $U$-integrations in the 
numerator in (\ref{exp2b}) \cite{F2}. 
Though a well-known fact (see e.g. \cite{R}), it is perhaps worth recalling here that within this hopping expansion there are no restrictions on how many times a bond is revisited in drawing all such possible connected graphs \cite{F3}.

This hopping expansion is naturally organized as an expansion in powers of $1/N_C$,  for fixed $N_F$. The trees graphs are all or order unity, whereas loops graphs contribute inverse powers of $N_C$. Note that, given this expansion of $G(x,x)$,  
the vev of $\bpsi\psi$ contains an extra overall multiplicative factor of $N_CN_F$ from the trace in (\ref{exp1}). 

The trees then give the leading contribution in the large $N_C$ limit. Self-consistent summation of all trees leads to the well known result \cite{BBEG} - \cite{KS}, \cite{ETT1} of  non-vanishing vev (\ref{exp1}) in the limit of vanishing external source ($m\to 0$), i.e. chiral symmetry breaking at strong coupling. 

The loop contributions provide non-leading small corrections for large $N_C$. They, however, also contain $N_F$ dependence. This can lead to drastically different state of affairs if 
$N_F$ is no longer taken to be small compared to $N_C$. 
Thus, among the graphs in Fig. \ref{csscF1} (b) the first is of order 
$(N_F/ N_C)$, the second is of order $(N_F/N_C)^2$, and the third is of order $(N_F/N_C)^4$. 
They are indeed subdominant for large enough $N_C$ at fixed $N_F$; but they become dominant in the opposite limit of large $(N_F/N_C)$. To consider this latter limit then we have to deal with the loop graphs.

\begin{figure}[ht]
\begin{center}
\includegraphics[width=0.95\columnwidth]{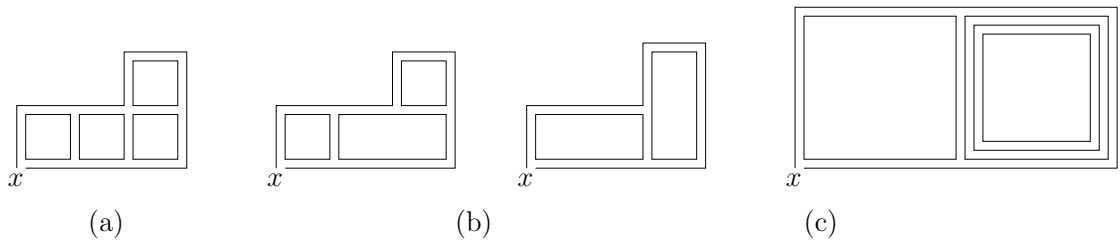}
\caption{(a) A maximally tiled graph of area $4$ attached at site $x$; (b) two non-maximal 
tilings of the graph; (c) a maximally tiled $2$-plaquette graph with one plaquette multiply tiled. 
(Plaquettes in (c) drawn to larger scale for clarity. Also, for simplicity, here and in subsequent figures we no longer always indicate the $U$ and $U^\dagger$ directions along each graph bond.) \label{csscF2}}
\end{center}
\end{figure}

Consider the $(N_F/N_C)$ order of graphs of a given non-zero lattice area $A$ (in plaquette units). 
The graphs of the highest possible order,  $(N_F/N_C)^A$, are those maximally tiled, i.e., tiled by 
a number of single plaquette loops equal to the area enclosed by the graph. Thus, the graph in Fig. \ref{csscF2}(a) is or order $(N_F/N_C)^4$; whereas in Fig. \ref{csscF2} (b) the first graph is of order $(N_F/N_C)^3$ and the second of order $(N_F/N_C)^2$. This rule holds whether a graph is planar or not (self-intersecting). If a plaquette is covered 
$k$ times by the graph folding onto itself, it contributes $k$ units to the area. The same applies in the case of multiply tiled plaquettes; a $k$-times tiled plaquette is also counted as contributing to the graph area $k$ units \cite{F4}. So, e.g., in Fig. \ref{csscF2} (c) the connected contribution from the 3-tiled plaquette results in order $(N_F/N_C)^4$ for the graph, which is then counted as being of area 4.

\begin{figure}[ht]
\begin{center}
\includegraphics[width=0.95\columnwidth]{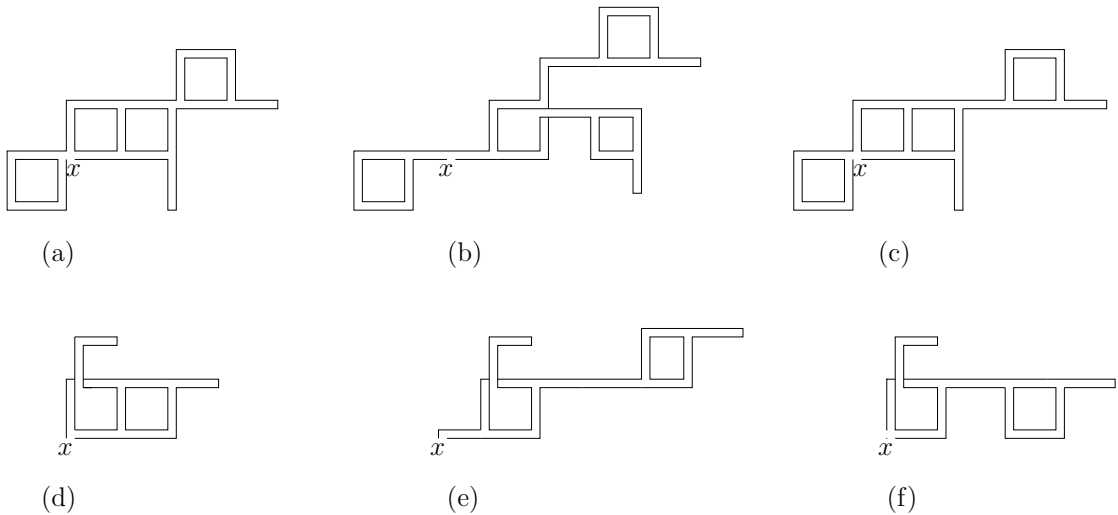}
\caption{An unreduced graph (a), and one reduced graph (b) and one partially reduced graph (c) 
arising from it; similarly for the unreduced graph (d) and reduced (e) and partially reduced (f) graphs. \label{csscF3}}
\end{center}
\end{figure}

We now classify the connected maximally tiled graphs as follows. 
Unreduced graphs are defined as those consisting of a set of  
tiled plaquettes where each plaquette is connected to at least one other plaquette in the set at a common boundary site or along a common boundary bond, or is attached to site $x$. Furthermore, on any plaquette or at site $x$, tree segments of any length and not connecting to any other plaquette may be attached. Thus all three graphs in Fig. \ref{csscF1} (b) belong to this class. 
Reduced graphs are those where each tiled plaquette in the graph is 
connected to one or more of the other plaquettes in the graph or to the site $x$ {\it only} by a tree segment, with any two plaquettes connected by at most one tree segment. 
Furthermore, on any plaquette or at site $x$, tree segments of any length and not connecting to any other plaquette may be attached. Thus, in common physics jargon, reduced graphs are 1PR with respect to any tree segment \cite{F5}.
Partially reduced graphs then are those containing one or more unreduced plaquette subsets connected to the rest of the graph or the site $x$ by tree segments. 
Examples are shown in Fig. \ref{csscF3}. 

Consider all unreduced graphs attached to site $x$ of given total area $A$ and having $k$ tree segments of total length $l$ attached
(cf. Fig. \ref{csscF3} (a), (d)). 
Any unreduced graph can be constructed by attaching one plaquette at a time either to the site $x$ or to another plaquette at a site or along a bond on its boundary. Now, counting all possible orientations, there are $2d(d-1)$ ways of attaching one tiled plaquette at site $x$. Each subsequent tiled plaquette can then be attached in $16(d-1)$ ways along a bond, and less than $32d(d-1)$ ways at a common boundary site, counting all possible positions and flips in every 2-plane. 
There are $1+8$ ways of attaching the first unit of each segment on a plaquette or at $x$, i.e, $9^k$ ways for 
$k$ segments; and since $1\leq k\leq l$, as each segment must be at least one bond long, the total number of ways of attaching total-length-$l$ tree segments is less than $2\times 9^l$. Furthermore, 
there are $(2d-1)$ ways of orienting each segment unit with a backtracking constraint (see below - this though does not really affect the argument).
So, for  the total number of unreduced graphs of area $A$ with tree segment of total length $l$ one has 
\beq 
n_u(A,l) <  4d(d-1) [16(d-1)(2d+1)]^{A-1} 9^l (2d-1)^l \,. \label{nubound} 
\eeq

Consider now the class of reduced graphs obtained by separating plaquettes by inserting 
tree segments (cf. Fig. \ref{csscF3} (b), (e)). 
A tiled plaquette can be attached to a tree segment endpoint in 
$8d(d-1)$ different ways. Each unit of a connecting segment can be oriented in $(2d-1)$ ways, so if the connecting segments have total length $l_c$, there are $(2d-1)^{l_c}$ ways of attaching them. Note that there must be $A$ such connecting segments between the plaquettes and $x$, so that $l_c\geq A$. In addition, $k$ further tree segments of total length $l$ can be attached to the $A$ plaquettes or the site $x$ in 
$(1+8)^k$ ways. Again, since $1\leq k\leq l$, the number of ways of attaching total-length-$l$ tree segments is greater than $9^l$, and they can be 
oriented in $(2d-1)^l$ ways. Actually, the first leg of segments attached directly at $x$ have no backtracking constraint (see below). So the total number of reduced graphs 
of area $A$, with connecting segments total length $l_c$ and other tree segments of total length 
$l$ attached is not less than ($d \geq 3$): 
\bea n_r(A, l_c, l) &\geq & [8d(d-1)]^A (2d-1)^{l_c} 9^l (2d-1)^l  \\
& \geq & 8d(d-1)(2d-1) [8d(d-1)(2d-1)]^{A-1}  9^l (2d-1)^l  \nonumber \\
& > & n_u(A,l) \;.  \nonumber \label{nrbound}
\eea 
By the same counting, for the number $n_{pr}(A, q, l_c , l)$ of partially reduced graphs 
with $q$ reduced plaquettes, $1\leq q <A$, and connecting segments of total length $(l_c-A+q)$, one has 
\beq n_u(A,l) < n_{pr}(A,q , l_c, l) < n_r(A,l_c, ,l) 
\,. \label{nprbound}
\eeq

The reduced graphs then dominate in number among maximally tiled graphs. They also clearly contain the highest powers of $1/m$.
The explicit evaluation of reduced graphs within the hopping expansion is straightforward and shows that each reduced tiled plaquette may be taken to contribute a factor of $(-1)(N_F/N_C)(1/4m^2)^4$,    
each tree segment unit (one bond long, double line) a factor of $(-1/4m^2)$,  
and the attachment to site $x$ a final factor of $(1/m)$. The 
value $I_{\gamma_r}$ of a reduced graph $\gamma_r$ of 
given $A$, $l_c$ and $l$ is then independent of how these building blocks are arranged, i.e. of the actual reduced graph structure:
\beq 
I_{\gamma_r}= (-1)^{A+l_c+l} \left({N_F\over N_C}\right)^A {1\over m} \left({1\over 4m^2}\right)^{4A+ l_c+l} \equiv I_r \,.  \label{redgraph}
\eeq
Unreduced ($q=0$) graphs miss at least a factor of $(-1/4m^2)^{l_c}$; this is for unreduced plaquettes 
attached at common boundary sites - unreduced plaquettes attached at common boundary bonds miss additional factors of $(1/2m)^2$. Partially reduced graphs
miss at least a factor $(-1/4m^2)^{A-q}$. The value $I_\gamma(A,q,l_c,l)$ of any unreduced or partially reduced graph $\gamma$ can then be expressed in terms of the fully 
reduced graph value (\ref{redgraph}) as 
\beq
 I_\gamma = c_\gamma (-4m^2)^{A-q} ( 4m^2)^{p_\gamma} I_r \;, \label{graphvalue} 
 \eeq
where $c_\gamma$ is a $\gamma$-dependent positive or negative numerical constant of order unity, and $p_\gamma\geq 0$ a $\gamma$-dependent non-negative integer power.  
Let $\Gamma_r(A,l_c,l)$  and $\Gamma_{pr}(A. q, l_c, l)$ 
denote the sets of reduced and of partially reduced or unreduced graphs, respectively, of given $A, l_c, l$. Then by (\ref{graphvalue}) one has 
\beq 
\sum_{q=0}^{A-1}\, \sum_{\gamma\in \Gamma_{pr}(A,q, l_c,l)} \!\!I_\gamma 
= 4m^2\, \R(A,l_c,l) \left( \sum_{\gamma\in \Gamma_r(A,l_c,l)} I_\gamma \right) \label{prsum1}
\eeq
with 
\beq  
\R(A,l_c,l) \equiv [n_r(A,l_c,l)]^{-1} \, \sum_{q=0}^{A-1} \sum_{\gamma\in \Gamma_{pr}(A,q, l_c,l)} \!\! c_\gamma (-4m^2)^{A-1-q} ( 4m^2)^{p_\gamma} \label{prsum2} \;.
\eeq 
For $m<1/2$ and any constant $c> c_\gamma$,  use of (\ref{nprbound}) gives the bound: 
\beq 
|\R(A,l_c,l) | < c \left[{1- (4m^2)^{A}\over 1-(4m)^2 }\right]   \label{prsum3}\,.
\eeq 

Consider then the contribution of all maximally tiled graphs up to a maximum size $A\leq K$, $\l_c\leq K$, $l\leq K$: 
\bea
& & \left(\sum_{A,l_c,l}^K \,
 \sum_{\gamma\in \Gamma_r(A,l_c, l)}\!\! I_\gamma\right)  + 
\left( \sum_{A,l_c,l }^K \,\sum_{q=0}^{A-1}\sum_{\gamma\in \Gamma_{pr}(A,q, l_c,l)}\!\! I_\gamma \right)  \nonumber \\
& = & \left( \sum_{A,l_c,l}^K \; \sum_{\gamma\in \Gamma_r(A,l_c, l)}\!\! I_\gamma \right) \Big[ 1 + \lambda(K,m)\Big]  \;. 
\label{leadsum} 
\eea
Using (\ref{redgraph}) and (\ref{prsum1}) one now obtains for the ratio of the contribution of all partially reduced graphs to that of all reduced graphs: 
\beq 
\lambda(K,m) = \left(\sum_{A,l_c,l}^K        \sum_{q=0}^{A-1}\sum_{\gamma\in \Gamma_{pr}(A,q, l_c,l)}\! \!I_\gamma \right)\,  \left( \sum_{A,l_c,l}^K \sum_{\gamma\in \Gamma_r(A,l_c, l)}\! \! I_\gamma \right)^{-1} 
= {\N(K)\over {\cal D}(K) }   \, , \label{ratiosum1}
\eeq
where 
\beq 
{\cal D} = 1 + \sum_{{\bf k} = (k_1,k_2,k_3) \atop k_1\geq k_2\geq1, k_3 \geq 1}^{K-1} 
 a_{\bf k} \bar{n}_{{\bf k},K}  \label{ratiosum2}
\eeq
\beq 
\N = 4m^2\, \left[ \R(K,K,K)  + \sum_{{\bf k} = (k_1,k_2,k_3) \atop k_1\geq k_2\geq1, k_3 \geq 1}^{K-1} 
 a_{\bf k}\, \bar{n}_{{\bf k},K}\, \R(K-k_1,K-k_2,K-k_3) \right]\label{ratiosum3}
 \eeq
 with 
 \beq
 a_{\bf k}= (-1)^{k_1+k_2+k_3} (-4m^2)^{4k_1+k_2+k_3} \left({N_C\over N_F}\right)^{k_1}   \label{ratiosum4}
 \eeq
 and 
 \beq 
 \bar{n}_{{\bf k},K}= { n_r(K-k_1,K-k_2,K-k_3)\over n_r(K,K,K) }
  < C_1^{-k_1} C_2^{-k_2} C_2^{-k_3} \label{ratiosum5} \;
  \eeq
for some constants $C_1$, $C_2$, $C_3$. 
The inequality in (\ref{ratiosum5}) follows easily from our counting estimates (\ref{nubound}) - 
(\ref{nprbound}) above. 
It follows from these expressions that the ratio (\ref{ratiosum1}) has a well-defined limit 
$K\to \infty$ for small $m$; indeed, by (\ref{prsum3}), one has  
\beq
\lim_{K\to \infty} |\lambda(K,m)| \leq \mbox{(constant)} \; m^2  \;.  \label{ratiosum6}
\eeq

We  have considered the 
maximally tiled graph contributions. It now should not be hard to see that counting and adding the contribution of the non-leading non-maximally tiled partially reduced graphs in (\ref{leadsum}) would only modify (\ref{ratiosum1}) by additional terms carrying 
additional $(N_C/N_F)$ suppression factors and higher $m$ powers, and thus not qualitatively alter this conclusion. 

It follows from (\ref{ratiosum6}) that if we are interested in the small $m$ limit, we need only consider the round bracket factor in (\ref{leadsum}), i.e., all reduced graphs. In other words,  
the expression for the propagator we need consider is: 
\beq 
G_{K}(x,x;m)\equiv  
\left( \sum_{A,l_c,l \atop l_c\geq A\geq 1, A\geq 0, l\geq 0}^K \; \sum_{\gamma\in \Gamma_r(A,l_c, l)}\!\! I_\gamma \right) \,  ,\label{leadsumred}
\eeq
were now we also include all  
pure tree graphs, which are reduced graphs by definition. 

For large enough $m$ the sums in (\ref{leadsumred}) are absolutely convergent  in the $K\to \infty$ and large volume limit. This is the statement of the hopping expansion convergence at large $m$. (The proof, by standard arguments, is actually already implicit in our graph counting above.)
To continue (\ref{leadsumred}) outside this region requires some resummation procedure. 
The sum (\ref{leadsumred}) can be performed by a self-consistent 
summation of all reduced graphs (directly in the $K\to \infty$ limit) leading to a self-consistent equation for $G(x,x)$ 
(section \ref{resumeq}). The solution to this equation for 
large $m$ reproduces, of course, the hopping graph expansion of (\ref{leadsumred}). 
But the equation obtained by the self-consistent graph summation is meaningful for all 
$m$, and one may, in particular, examine whether any physical solutions exist as $m\to 0$. 


\section{Self-consistent graph summation and equation for $G(x,x)$ \label{resumeq}}
\begin{figure}[ht]
\begin{center}
\includegraphics[width=0.6\columnwidth]{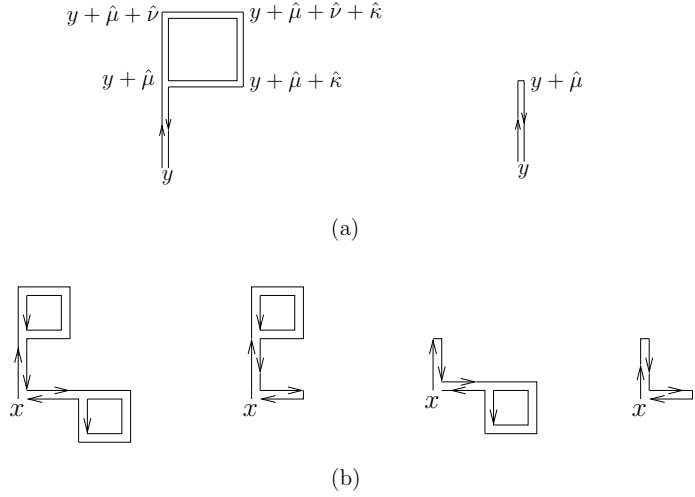}
\caption{(a) The two basic structures attached to a site $y$ whose iteration generates all reduced graphs. (b) Attachment of the two basic structures in (a) $n=2$ times to the site $x$ (two particular 
directions are shown) to begin generating all ``two-trunk" reduced graphs. 
\label{csscF4}}
\end{center}
\end{figure}

All reduced graphs can be generated by iteration of the two basic structures shown in Fig. \ref{csscF4} (a). Summation over all possible orientations for both the trunks and the attached plaquette in Fig. \ref{csscF4} (a) is always implied when attaching these two basic structures at any site. The simplest contributions are ``one-trunk" graphs obtained by attaching these two basic structures at the given site $x$.
Next, there are contributions from $n$-trunk graphs, any integer $n$, obtained by attaching $n$ of the two basic structures  at $x$.  Fig. \ref{csscF4} (b) shows this for $n=2$. The complete set of reduced graphs can now be generated in a recursive manner by attaching these structures again at each available site $x+\hmu$, $x+\hmu+\hnu$, $x+\hmu+\hnu +\hkappa$, $x+\hmu+\hkappa$ on top of each trunk of these $n$-trunk graphs. Indeed, this amounts to starting generating the complete propagator $G(y,y)$ at $y$ again, where $y$ stands for any of these sites. Note that the subset of {\it pure} trees is generated \cite{BBEG} by iteration of the second basic structure of Fig. \ref{csscF4} (a) above by itself. 

\begin{figure}[ht]
\begin{center}
\includegraphics[width=0.6\columnwidth]{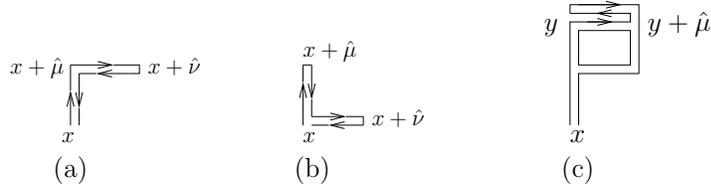}
\caption{2-trunk tree at site $x$ in (b) is contained in the set of 1-trunk trees at $x$ in (a) if all $2d$ directions are allowed at $x+\hat{\mu}$, i.e., backtracking is allowed. (c) Similar ambiguity for the attachment of a unit segment at $y$ versus $y+\hat{\mu}$ on top of the first basic structure in Fig. 
\ref{csscF4} (a). 
\label{csscF5}}
\end{center}
\end{figure}

In this recursive rebuilding, however, there can be a slight overcounting of tree segments \cite{MS},  \cite{ETT1} as illustrated in Fig. \ref{csscF5}. Though inessential for the conclusion, this overcounting is easily corrected by reducing by one (no backtracking) the available lattice directions when attaching tree segments at sites beyond the initial point $x$. This is done by performing the recursive rebuilding in two steps \cite{MS}, \cite{ETT1}: one (i)
first recursively builds a complete propagator $G_I(y,y)$ at site $y$ incorporating the no-backtracking constraint on tree segments; then (ii) obtains the complete propagator $G(x,x)$ 
by attaching $G_I$ on top of the $n$-trunk basic structures attached at $x$. 

\begin{figure}[ht]
\begin{center}
\includegraphics[width=0.8\columnwidth]{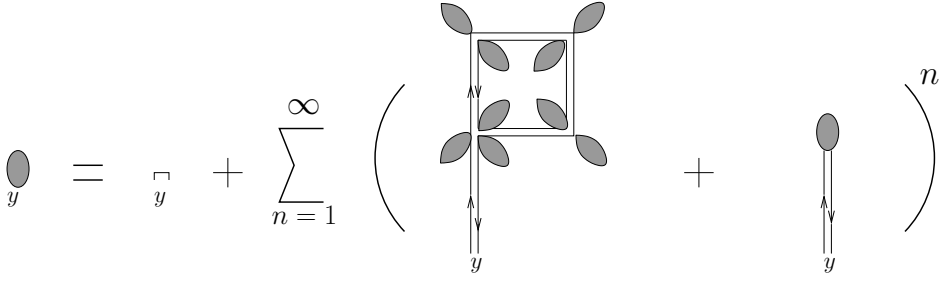}
\caption{Self-consistent equation for $G_I$ (cf. eq. (\ref{scGI})). The hatched blob represent the full $G_I$. 
\label{csscF6}}
\end{center}
\end{figure}

This recursive building of all reduced graphs, subject to the no-backtracking constraint, then immediately implies the self-consistency relation for $G_I$ graphically depicted in 
Fig. \ref{csscF6}. The first ($n=0$) term stands for the lowest order contribution given by just the bare propagator (\ref{bareG}). 
Note that, with space-time independent source $m$, $G_I(y,y)=G_I$, and similarly $G(y,y)=G$, are in fact site-independent by translation invariance. 
Letting $G_I= g(m) {\bf 1_{\ssc C}}{\bf 1_{\ssc F}}$, and using 
the explicit expressions (\ref{act3a}) and (\ref{bareG}) the equation in Fig. \ref{csscF6} becomes: 
\bea
g(m) & = &  \left[ 1 + \sum_{n=0}^\infty \left[{1\over m} \left(4d^2(d-1)\left({N_F\over N_C}\right) \left({g(m)\over2}\right)^9 - {(2d-1)\over2}\left({g(m)\over 2}\right) \right)\right]^n \right]
{1\over m} \nonumber   \\
& = & \left[ m -  \left(4d^2(d-1)\left({N_F\over N_C}\right) \left({g(m)\over2}\right)^9 - {(2d-1)\over2}\left({g(m)\over 2}\right) \right)\right]^{-1} \; . \label{scGI} 
\eea 
(\ref{scGI}) is to be solved for $g(m)$. The complete propagator $G$ is then given in terms of 
$G_I$ by:
\beq
G = 
\left[ m -  \left(4d^2(d-1)\left({N_F\over N_C}\right) \left({g(m)\over2}\right)^9 - {d}\left({g(m)\over 2}\right) \right)\right]^{-1}  {\bf 1_{\ssc C}}{\bf 1_{\ssc F}} \; . \label{G} 
\eeq

Had we allowed the slight graph overcounting avoided through  (\ref{scGI}) - (\ref{G}), we would have obtained a self-consistent equation directly for $G$ by now writing $G=g(m){\bf 1_{\ssc C}}{\bf 1_{\ssc F}}$ with $g(m)$ satisfying: 
\beq
g(m) = \left[ m -  \left(4d^2(d-1)\left({N_F\over N_C}\right) \left({g(m)\over2}\right)^9 - d\left({g(m)\over 2}\right) \right)\right]^{-1}  \; . \label{scG} 
\eeq
Clearly, existence of a solution for 
(\ref{scGI}) implies the same for (\ref{G}) and vice-versa.

\section{Absence of chiral symmetry breaking at large $N_F/N_C$ \label{noCSB}}
We consider then solutions to (\ref{scGI}). Letting $z=g/2$ it assumes the form: 
\beq
F(z,m) + 1 = 0  \, , \label{scGIa}
\eeq 
where 
\beq 
F(z,m) \equiv 8d^2(d-1) \left({N_F\over N_C}\right) z^{10}  - (2d-1) z^2    - 2m z  \,. 
\label{Fdef} 
\eeq
For sufficiently large $m$ at fixed  $(N_F/N_C)$
(\ref{scGIa}) is immediately seen to have a solution of the form 
\beq
 2z = g(m) = 1/m + O(1/m^3) \;. \label{largemsoln}
\eeq
This, as noted above,  is indeed the expected large $m$ hopping expansion solution. 

We are, however, interested in solutions for small $m$.  
Now, 
$F(0,0)=0$ and $F(z,0) >0$ for $|z| > [(2d-1)/8d^2(d-1)]^{1/8} (N_C/N_F)^{1/8}$;  whereas at its
minimum $z_{\rm m}= \pm [(2d-1)N_C/ 40d^2(d-1)N_F]^{1/8}$ this function has the value 
\beq 
F(z_{\rm m},0) = - {4\over 5} \left[{(2d-1)^5 \over  40d^2(d-1)}\right]^{1/4} \left({N_C\over N_F}\right)^{1/4}
  \;. \label{Fmin}
\eeq
It follows that $F(z,0) + 1 > 0$  for all $z$, and (\ref{scGIa}) has no real solutions, provided 
\beq 
 \left({4\over 5}\right)^4 \left[{(2d-1)^5 \over  40d^2(d-1)}\right]  <  \left({N_F\over N_C}\right)   \;. \label{Fnobound}
\eeq
For $(N_F/N_C)$ values satisfying (\ref{Fnobound}) then there is no real solution in the absence of sources, i.e. $m=0$, and no condensate forms.  Equality in (\ref{Fnobound}) gives an estimate of 
the critical $(N_F/N_C)_c$ at $m=0$. 

Similarly, for $m >0$ there is a critical $(N_F/N_C)_c$, depending on $m$, above which there is no real solution to (\ref{scGIa}). 
It is immediately seen from (\ref{scGIa}) - (\ref{Fdef}) that this critical $(N_F/N_C)_c$ is an increasing function of $m$. 
Thus, for a given $(N_F/N_C)$ above the bound (\ref{Fnobound}) there is 
a mass $m$ for which this $(N_F/N_C)$ is critical. 
A real solution exists for $m^\prime > m$, 
which approaches (\ref{largemsoln}) for $m^\prime$ large enough, but it turns complex when $m^\prime < m$.

In contrast, for $(N_F/ N_C)$ violating the bound (\ref{Fnobound}), (\ref{scGIa}) has real solutions for all $m$. 
In particular, in the $(N_F/ N_C)\to 0$ limit, (\ref{scGIa}) is solved at 
$m=0$ by $g(0) = 2/\sqrt{2d-1}$, which, substituted in (\ref{G}), gives  
\beq 
G =  {2\over d \, g(0)} {\bf 1_{\ssc C}}{\bf 1_{\ssc F}} = 
\sqrt{2\over d}\,\sqrt{1- {1\over 2d}}\;  {\bf 1_{\ssc C}}{\bf 1_{\ssc F}}\;,  \label{csb}
\eeq
which in turn, by (\ref{exp1}), yields the well-known result for non-vanishing chiral condensate 
at large $(N_C/N_F)$ \cite{BBEG} - \cite{KS}, \cite{ETT1}. For small $m$ the solution \cite{F7} gives a condensate proportional to $m$, whereas at large $m$ it approaches (\ref{largemsoln}). 

It is important to note that this $m=0$ real solution at small $(N_F/N_C)$ does not drop to zero as $(N_F/N_C)$ is increased to the  critical value. 
Rather, at the critical value, the solution turns complex signaling that a condensate 
ceases to form. This indicates the presence of a first order transition,  as indeed seen in the Monte-Carlo data \cite{deFetal}. 
The feature that $(N_F/N_C)_c$ is an increasing function of $m$ is also clearly seen in these data.  
The bound (\ref{Fnobound}) gives $N_F \geq  11$ for $N_C=3$ in good agreement with 
$N_F \simeq 13$ staggered flavors found in \cite{deFetal} - with baryon tilings, which  are generically suppressed by additional factors of $1/N_C$ and were omitted here, providing some additional small correction. 
In this connection  also note that the method used here, i.e., expansion at large $m$, resummation 
and continuation in $m$, can approach  
this phase transition boundary only from the chirally broken phase. 
As always for two such distinct phases, a different type of expansion would be required to 
directly describe the other, i.e., the chirally symmetric phase. 
Finally, it is worth noting that, regardless of other specific numerical details,  
the structure of the resulting resummation equation (\ref{scGIa}), (\ref{Fdef}) reflects 
the robustness of this chiral symmetry restoration effect produced by sufficient number of light fermion flavors, i.e. large enough $(N_F/N_C)$.

\section{Conclusion \label{Con}} 

We have considered $U(N_C)$ and $SU(N_C)$ lattice gauge theories at strong coupling ($\beta=0$) coupled to $N_F$ flavors of staggered fermions and investigated the formation of a chiral symmetry breaking condensate within a fermionic hopping expansion. Analysis and classification of graphs in their dependence on $N_C$, $N_F$ and $m$ led us through a graph resummation procedure to an equation for the determination of the condensate. At low values of the parameter $(N_F/N_C)$ this yields the well-known solution for a non-vanishing condensate persisting at $m=0$. 
At large values of $(N_F/N_C)$, however, different classes of graphs dominate and the situation changes drastically: there is a critical value of $(N_F/N_C)$ above which no real solution for a condensate exists. The sudden disappearance of a non-vanishing condensate solution at a 
critical $(N_F/N_C)$ indicates a first order transition. 
This result, as well as some other qualitative features of behavior as a function of $(N_F/N_C)$ and $m$ are in agreement with the recent simulation results for $N_C=3$ in \cite{deFetal}. 

There are some immediate obvious extensions of this work one could consider. 
Including the baryon sector in $SU(N_C)$ in the 
hopping expansion can be done, but it will not materially alter the conclusion. 
The effect of non-vanishing but small plaquette coupling can be incorporated by strong coupling expansion in $\beta$. 
It is straightforward to incorporate the corrections due to this expansion, which is convergent for small enough $\beta$, in our formalism above. It is in fact easy to see that doing this has one qualitative effect: it lowers the 
value of the critical $(N_F/N_C)$ as a function of increasing $\beta$. 
It would be interesting to investigate how far this can be pushed into the higher $\beta$ region by analogous graph resummations and/or Pad\'{e} approximants, and whether new phase boundaries appear in the $(\beta, (N_F/N_C), m)$-space.  

The remarkable feature here is of course the existence of  strongly coupled lattice gauge systems with 
colorless ("confined") excitation spectrum and chiral symmetry. This clearly opens up   
several issues, in particular the basic field-theoretic question of the complete phase 
diagram in  $(\beta, (N_F/N_C), m)$ and the issue of the existence of continuum limits \cite{F8}. 
More generally, such gauge systems invite consideration 
of theories strongly coupled in the UV, possibly controlled by a non-trivial UV fixed point, and with non-trivial or trivial IR fixed points. 
There are also several areas of potential phenomenological application. One may indeed imagine a gauge theory defined at some short scale on a lattice, either physical or serving as a cutoff. In strongly correlated electron systems such gauge systems often naturally emerge as effective descriptions of underlying electron interactions. Broken chiral symmetry in the gauge system would correspond in this case to an anti-ferromagnetic phase. There have been suggestions that a chiral symmetry preserving gauge theory, on the other hand, would describe the (high frequency regime of) the pseudo-gap phase in the 
high $T_c$ superconductivity phase diagram \cite{GW}. 
In particle physics such models  
offer a mechanism for obtaining light bound states which suggests some potential phenomenological applications to which we hope to turn elsewhere. 

\vspace{1.5cm} 

The author would like to acknowledge the hospitality and support of KITPC, Beijing, where this work was initiated. He would like to thank the participants of the program ``Critical behavior of lattice models" (July-August, 2012), and in particular Xi Cheng and Xiao-Gang Wen for discussions that motivated this work, and P. de Forcrand for correspondence.  
This research was also partially supported by NSF-PHY-0852438.


\begin{thebibliography}{99}
\bibitem{F0} The expansion amounts to a loop expansion with a modified gauge boson propagator which includes the fermion bubble contribution to the self-energy, and modified $n$-point gauge boson vertices which include the contribution of the fermion loop with $n$ external bosons; 
and all gauge boson interaction legs carrying $1/\sqrt{N_F}$ factors. The expansion correctly indicates that the theory should become more ``semiclassical" for large number of fermion flavors, but clearly cannot deal with the strong coupling limit.
\bibitem{F1} More generally, the global symmetry is $G\times G$ where 
$G$ is the subgroup of $U(N_F)$ which commutes with all the matrices $\{ \gamma[b]\}$. 
For staggered fermions $G=U(N_F)$.
\bibitem{F2} Closed paths not so coupled give of course disconnected graphs canceling against the denominator.
\bibitem{F3} Indeed, the rules of Grassmann integration automatically take into account the exclusion principle summing over all fermionic path integral configurations so as to give the appropriate cancellations 
between configurations contributing to connected and disconnected graphs. As a result, after the factoring of disconnected graphs, one is left with unrestricted sums for the remaining connected 
graphs \cite{R}.
\bibitem{F4} Note that the tilings must always conform to the rule of equal number of $U$'s and $U^\dagger$'s on each bond in order to obtain a non-vanishing result upon integration over the gauge field.
\bibitem{F5} Equivalently, viewing the plaquettes as vertices, a reduced graph is a tree.
\bibitem{F8} Some interesting proposals based on simulation results have been made in \cite{deFetal}. For earlier, well-known discussions, which, however, assume chiral symmetry breaking at sufficiently strong coupling for any number of flavors, see \cite{MY}. In this connection also note that a first order phase transition at fixed $N_F$, within the conformal window, to a chirally broken phase as $\beta$ is decreased has been seen in \cite{Hetal}. In the present context this might be interpreted as crossing the phase boundary $(N_F/N_C)_c(\beta)$ extending into the $(\beta, (N_F/N_C))$ plane from the point $\beta=0$.
\end{thebibliography}
\end{document}